\documentclass[aps,prl,twocolumn,superscriptaddress]{revtex4}

\usepackage{graphicx}% Include figure files
\usepackage{dcolumn}% Align table columns on decimal point
\usepackage{bm}% bold math
\usepackage{color}

\begin{document}
\title{Magnetic Field Induced Instabilities in Localised Two-Dimensional Electron Systems}
\author{M. Baenninger}
\email{matthias.baenninger@cantab.net}
\affiliation{Cavendish Laboratory, University
of Cambridge, J.J. Thomson Avenue, Cambridge CB3 0HE, United
Kingdom.}
\author{A. Ghosh}
\affiliation{Department of
Physics, Indian Institute of Science, Bangalore 560 012, India.}
\author{M. Pepper}
\affiliation{Cavendish Laboratory, University of Cambridge, J.J.
Thomson Avenue, Cambridge CB3 0HE, United Kingdom.}
\author{H. E. Beere}
\affiliation{Cavendish Laboratory, University of Cambridge, J.J.
Thomson Avenue, Cambridge CB3 0HE, United Kingdom.}
\author{I. Farrer}
\affiliation{Cavendish Laboratory, University of Cambridge, J.J.
Thomson Avenue, Cambridge CB3 0HE, United Kingdom.}
\author{D. A. Ritchie}
\affiliation{Cavendish Laboratory, University of Cambridge, J.J.
Thomson Avenue, Cambridge CB3 0HE, United Kingdom.}
\date{\today}

\begin{abstract}
We report density dependent instabilities in the localised regime of mesoscopic two-dimensional electron systems (2DES) with intermediate strength of background disorder. They are manifested by strong resistance oscillations induced by high perpendicular magnetic fields $B_{\perp}$. While the amplitude of the oscillations is strongly enhanced with increasing $B_{\perp}$, their position in density remains unaffected. The observation is accompanied by an unusual behaviour of the temperature dependence of resistance and activation energies. We suggest the interplay between a strongly interacting electron phase and the background disorder as a possible explanation.
\end{abstract}
\maketitle

The behaviour of electrons in presence of both disorder and interactions is one of the most fundamental questions in condensed matter physics~\cite{Lee1985}. While in strongly disordered and weakly interacting systems, single-particle localisation is expected~\cite{Mott1975}, the formation of a Wigner crystal (WC) has been predicted for the opposite regime of very strong interactions and absence of disorder~\cite{Wigner1934}. Between those two regimes a variety of states has been proposed, ranging from a glassy phase~\cite{Davies1982} to a pinned WC~\cite{Chui1995}. A difficulty in the experimental investigation of interaction effects in disordered systems is that long-range disorder can cause charge inhomogeneities which lead to a percolation type transport, masking possible interaction effects~\cite{Sarma2005}.

Recently, a new approach for studying the influence of short-range disorder on strongly interacting 2DES has lead to the observation of a low-temperature collapse of electron localisation~\cite{Baenninger2008} and an interaction dominated hopping magnetoresistance~\cite{Ghosh2004,Baenninger2005,Baenninger2008a}. These results are strong evidence of a breakdown of conventional single-particle transport generally observed in the strongly localised regime of 2DES~\cite{Mott1968,Efros1975}. The experiments were carried out in mesoscopic electron systems in modulation doped GaAs/AlGaAs heterojunctions, extending only over a few microns in order to reduce the impact of long range charge inhomogeneities. The strength of disorder was controlled by varying the spacer width between 2DES and charged dopants which are the main source of disorder in these system.
\begin{figure}[!h]
\centering
\includegraphics[width=0.495\textwidth]{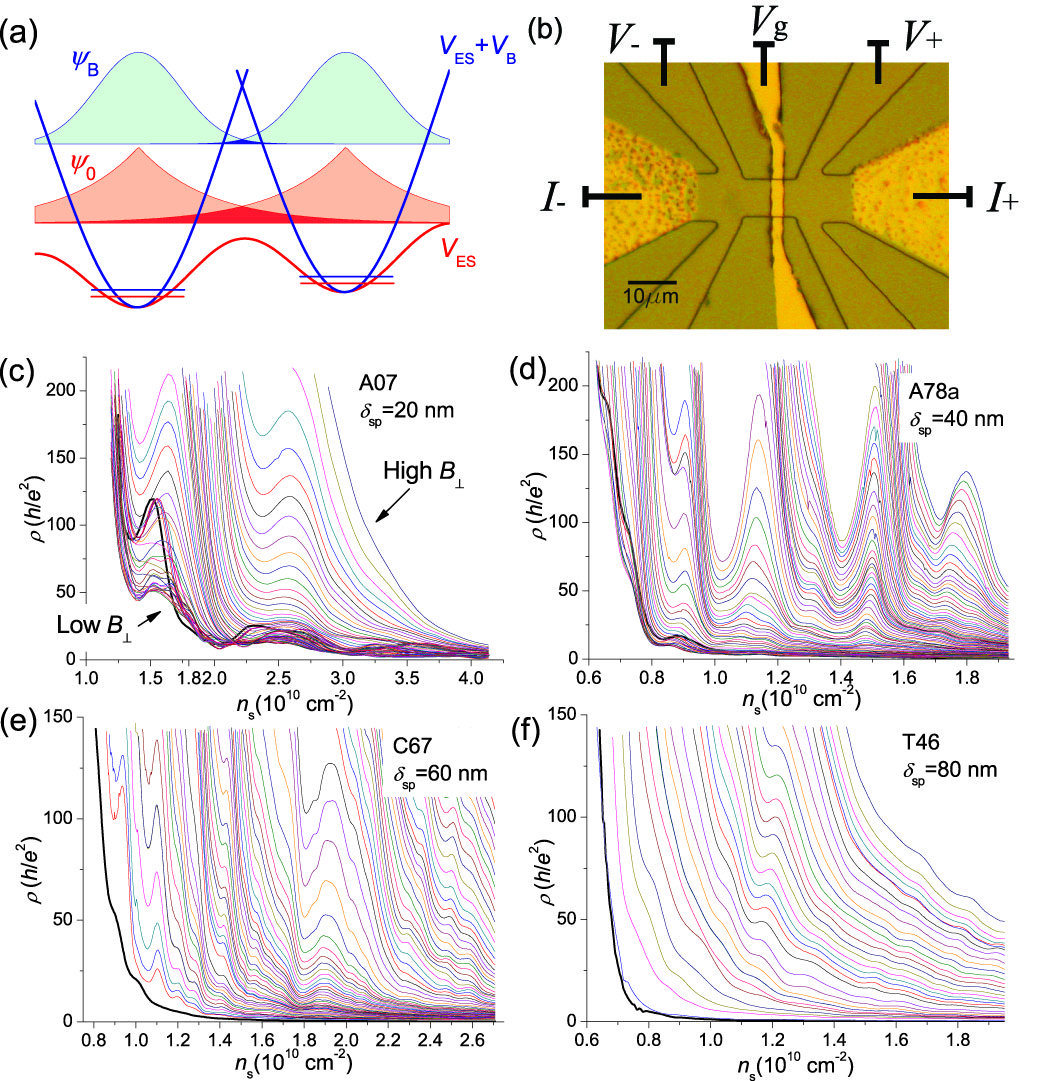}
\caption{(Colour online). (a) Schematic of two localised states in the minima of the self-consistent electrostatic potential at $B_{\perp}=0$ (red line) and at a finite field with additional magnetic confinement (blue line). The corresponding wave functions and their overlap are also indicated. (b) Microscope picture of a typical device with Hall bar mesa and topgate. Current and voltage probes for four-probe measurements are indicated. (c)-(f) Overview of magnetic field induced resistance oscillation for devices from four different wafers. At $B=0$ $\rho$ increases mostly monotonically with decreasing $n_{\rm{s}}$ (thick black lines), but a perpendicular magnetic field induces strong oscillations. (c) $B_{\perp}=0-3$\,T, $\Delta B_{\perp}=0.05$\,T. (d) $B_{\perp}=0-4$\,T, $\Delta B_{\perp}=0.05$\,T. (e) $B_{\perp}=0;$ $0.92-6.7$\,T, $\Delta B_{\perp}=0.08$\,T. (f) $B_{\perp}=0-9.6$\,T, $\Delta B_{\perp}=0.3$\,T.}
\label{fig1}
\end{figure}

A magnetic field perpendicular to a 2DES introduces an effective parabolic confinement potential, which can localise electrons, reducing the wave function overlap between them [see schematic in Fig.~\ref{fig1} (a)], and quench their kinetic energy. This may effectively increase the strength of their Coulomb interaction. In this Letter, we investigate the impact of strong magnetic fields and its interplay with disorder on transport in mesoscopic 2DES. All devices were made from Si $\delta$-doped heterojunctions with dimensions of the gated 2DES area $W\times L=8\,\mu\rm{m}\times 0.5-4\,\mu\rm{m}$ ($W$ width of the mesa, $L$ the length of the gate). A typical device is shown in Fig.~\ref{fig1} (b) and a list of devices is given in Table~\ref{table1}. Measurements were carried out in a $^3$He cryostat (base temperature 300\,mK) with standard lock-in four probe setup with excitation frequency $\Omega_{\rm{ext}}\sim 7$\,Hz and current $I_{\rm{ext}}$=100\,pA. The low current ensured linear response and minimal Joule heating. A slow cooldown from room temperature to 4\,K over $\gtrsim 24$\,h gave best results. This could be related to enhanced correlations between charges in the dopant layer, which may reduce the long range disorder fluctuations~\cite{Efros1988}. The gate voltage dependence of electron density $n_{\rm{s}}$ was measured with a technique based on the reflection of edge states in the Quantum Hall regime~\cite{Ghosh2004}.
\begin{table}[h]
%\centering
\caption{List of devices with relevant properties. $\delta_{\rm{sp}}$ and $d_{\rm{g}}$ are the spacer width and total depth of the 2DES, respectively and $W$ and $L$ the dimension of the gated area of the 2DES. $n_{\delta}$ is the ($\delta$)-doping concentration and $n_0$ and $\mu$ are as-grown electron density and mobility, respectively.\newline}
\begin{tabular}{c c c c c c c}
  \hline\hline\\[-8pt]
  % Device & Wafer & $\delta_{\rm{sp}} (nm) $ & $d_{\rm{g}}$ (nm) & $W\times L$ ($\mu\rm{m}\times\mu\rm{m}$) \\ [0.5ex]
  Device & $\delta_{\rm{sp}}$ & $d_{\rm{g}}$ & $W\times L$ & $n_{\delta}$ & $n_0$ & $\mu_0$\\
  & nm  &  nm  &  $\mu\rm{m}\times\mu\rm{m}$  &  $10^{12}$\,cm$^{-2}$  &  $10^{11}$\,cm$^{-2}$  &  cm$^2$/Vs\\ [0.5ex]
  \hline\\[-8pt]
  A07 & 20 & 120 & 8$\times$3 & 2.5 & 2.9 & $0.6\times 10^6$\\
  A78a & 40 & 290 & 8$\times$2 & 2.5 & 2.1 & $1.8\times 10^6$\\
  A78b & 40 & 290 & 8$\times$4 & 2.5 & 2.1 & $1.8\times 10^6$\\
  C67 & 60 & 290 & 8$\times$3 & 0.7 & 1.0 & $1.2\times 10^6$\\
  T46 & 80 & 300 & 8$\times$3 & 1.9 & 0.8 & $0.9\times 10^6$\\ [0.5ex]
  \hline\hline
\end{tabular}
\label{table1}
\end{table}

Figs.~\ref{fig1} (c)-(f) show the resistivity $\rho$ as a function of $n_{\rm{s}}$ at stepwise increasing $B_{\perp}$ for four devices made from wafers with spacer widths $\delta_{\rm{sp}}$=20, 40, 60 and 80\,nm.
While at zero magnetic field the resistance increases with decreasing $n_{\rm{s}}$ in a relatively featureless way (thick black lines), when a magnetic field is applied, strong resistance oscillations appear with maximum amplitudes $\Delta\rho>100\,h/e^2\approx 2.5$\,M$\Omega$. The amplitudes get larger with increasing $B_{\perp}$, but the positions of peaks and troughs is virtually unaffected by changes of several Tesla. The positions were also highly stable over experimental runs of several weeks, including many gate voltage sweeps and $T$ dependence measurements up to $\sim4.5$\,K. The oscillations start developing at $\rho\gtrsim 10\,h/e^2$, which highlights the importance of the 2DES being localised. While peaks generally start appearing only at Landau level (LL) filling factor $\nu=h n_{\rm{s}}/e B_{\perp}\lesssim 0.5$, no systematic connection to a certain filling factor can be observed and $\nu$ often changes by a factor of $2-3$ as a peak evolves at constant $n_{\rm{s}}$. This clearly rules out effects related to LLs, where peak positions would have to move with $B_{\perp}$. The strength of the oscillations is similar for devices with $\delta_{\rm{sp}}=20-60$\,nm [Fig.~\ref{fig1} (c)-(e)], but they practically disappear in a wafer of very low disorder with $\delta_{\rm{sp}}=80$\,nm [Fig.~\ref{fig1} (f)]. Macroscopic devices ($W\times L=100\,\mu\rm{m}\times 900\,\mu\rm{m}$) did not show any magnetically induced resistance oscillations (not shown). 

The separation between two resistance peaks in terms of $n_{\rm{s}}$ is usually $\Delta n_{\rm{s}}\gtrsim 0.2\times 10^{10}$\,cm$^{-2}$, which corresponds to adding several hundred electrons to the active area of the 2DES for typical device dimensions. This makes it very unlikely that the observation is caused by single electron charging effects such as Coulomb blockade (CB) oscillations, which have previously been reported in mesoscopic 2DES~\cite{Cobden1999}. Equivalently, the spacing in terms of gate voltage $\Delta V_{\rm{g}}\approx10-20$\,mV e.g. for the device shown in Fig.~\ref{fig1} (d), would suggest a dot area $e/C\Delta V_{g}\approx(150-200\,\rm{nm})^2$ with $C=\epsilon\epsilon_0/d_{\rm{g}}$ ($d_{\rm{g}}$ the separation between gate and 2DES) the capacitance per area for a parallel plate capacitor. This is a tiny area compared to the overall device area. In stark contrast to Fig.~\ref{fig1} (c)-(e) is the data shown in Fig.~\ref{fig4} (a). It shows the resistivity as a function of gate voltage for device A78b after an uncontrolled fast cooldown from room temperature to 4\,K. Here, a non-monotonic $V_{\rm{g}}$ dependence of $\rho$ is observed even at zero field (think black line) and with stepwise changes of $B_{\perp}$, fast resistance oscillations appear and disappear, with positions of peaks and troughs changing in an apparently random manner. A very similar behaviour has also been reported in strongly disordered mesoscopic devices with $\delta_{\rm{sp}}=10$\,nm~\cite{Ghosh2004a}. A ``bad'' cooldown, or small spacer may lead to larger disorder and, hence, to strong inhomogeneity of the electron distribution even on a mesoscopic length scale. This could lead to transport by tunnelling between electron ``puddles'', where CB effects become important~\cite{Raikh1991,Tripathi2006}. The data shown in Fig.~\ref{fig1} (c)-(e) shows a qualitatively different behaviour. We suggest that the smoother background disorder in a ``good'' cooldown leads a homogeneous electron distribution on the scale of the device size. Indeed, in several controlled slow cooldowns, device A78b showed a very similar behaviour to A78a [Fig.~\ref{fig1} (d)]. Additional evidence for a qualitatively different phenomenon comes from the fact that a $\cosh^{-2}$ line shape expected for CB could be fitted satisfactorily to the oscillations observed in the ``bad'' cooldown as well as in the devices with $\delta_{\rm{sp}}=10$\,nm~\cite{Ghosh2004a}. However, for the oscillations shown in Fig.~\ref{fig1} (c)-(e), the agreement of the fit was poor. 
\begin{figure}[h]
\centering
\includegraphics[width=0.495\textwidth]{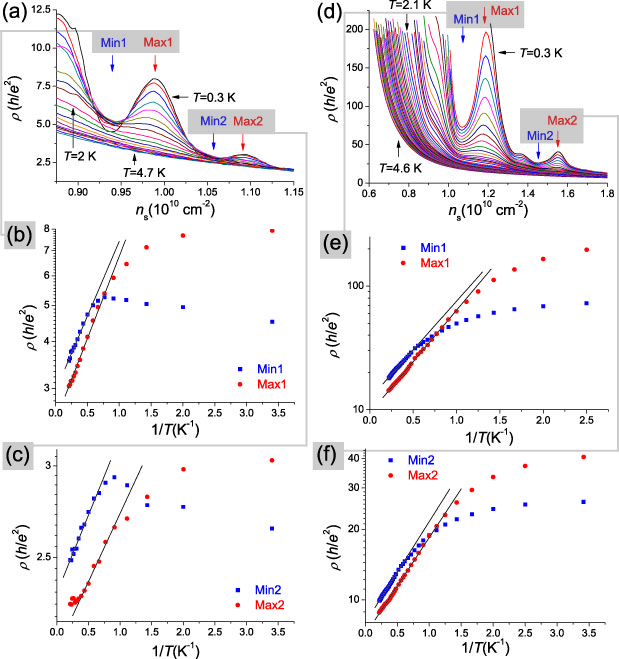}
\caption{(Colour online). $T$ dependence of resistance oscillations. (a)-(c) C67, $B_{\perp}$=0.5\,T. The high $T$ part is qualitatively the same at peaks and troughs in resistance and the activation energy [slopes of $\rho(T)=\rho_0 \exp(E_0/k_{\rm{B}}T)$ fits (solid lines)] is virtually unchanged between adjacent peaks and troughs. By contrast, the low $T$ behaviour changes qualitatively with the oscillations. At the peaks the behaviour remains weakly insulating, but the troughs show a metal-like $T$ dependence. (d)-(f) A78a, $B_{\perp}$=2\,T. The high $T$ behaviour is similar to (a)-(c), but at low $T$ no metal-like behaviour is observed in the troughs. However, a difference persists, with the saturation at the troughs being stronger and starting at higher $T$ than at peaks.}
\label{fig2}
\end{figure}

The $T$ dependence of $\rho$ in the regime of strong oscillations for two devices at $B_{\perp}$=0.5 and 2\,T is given in Fig.~\ref{fig2}. Panels (a) and (d) show the oscillations at stepwise changing $T=0.3-4.7$\,K. The oscillations are damped strongly with increasing $T$ and have virtually disappeared by $T\gtrsim 2$\,K. Panels (b), (c), (e) and (f) show $\rho$ vs. $1/T$ at several values of $n_{\rm{s}}$, each panel comparing the $T$ dependence of a resistance trough to the next peak with increasing $n_{\rm{s}}$. Generally, the $T$ dependence in our devices at low enough electron densities can be divided into two regimes: At high $T$, a strongly insulating (activated) behaviour is always observed, but at low enough temperatures a qualitatively different behaviour occurs with $\rho$ becoming weakly $T$ dependent (saturated) or even decreasing with decreasing temperature (metallic)~\cite{Baenninger2008,Ghosh2004,Baenninger2005,Baenninger2008a}. This finding is confirmed by the results in Fig.~\ref{fig2}. Let us first focus on the low $T$ part. The 60\,nm spacer device at relatively small field $B_{\perp}=0.5$\,T shows a striking difference between peaks and troughs [Fig.~\ref{fig2} (a)-(c)]. At the positions of peaks, the $T$ dependence weakens at lowest temperatures and deviates from the high $T$ activated behaviour. The troughs also show such a deviation, however, here, the $T$ dependence is non-monotonic and $\rho$ starts dropping as $T$ is lowered further. This observation clearly suggests a connection between the origin of the oscillations and the previously reported metallic behaviour in our devices~\cite{Baenninger2008}. At a higher field $B_{\perp}=2$\,T for the 40\,nm spacer device a clear difference between peaks and troughs is still observed [Fig.~\ref{fig2} (d)-(f)]. However, the metallic behaviour is replaced with a strong saturation. This is again in agreement with previous observation that the metallic phase was suppressed at $B_{\perp}\gtrsim 1-1.5$\,T~\cite{Baenninger2008}. The saturation is much stronger in case of troughs than peaks and the deviation from exponential behaviour sets in at higher $T$.

\begin{figure}[h]
\centering
\includegraphics[width=0.3\textwidth]{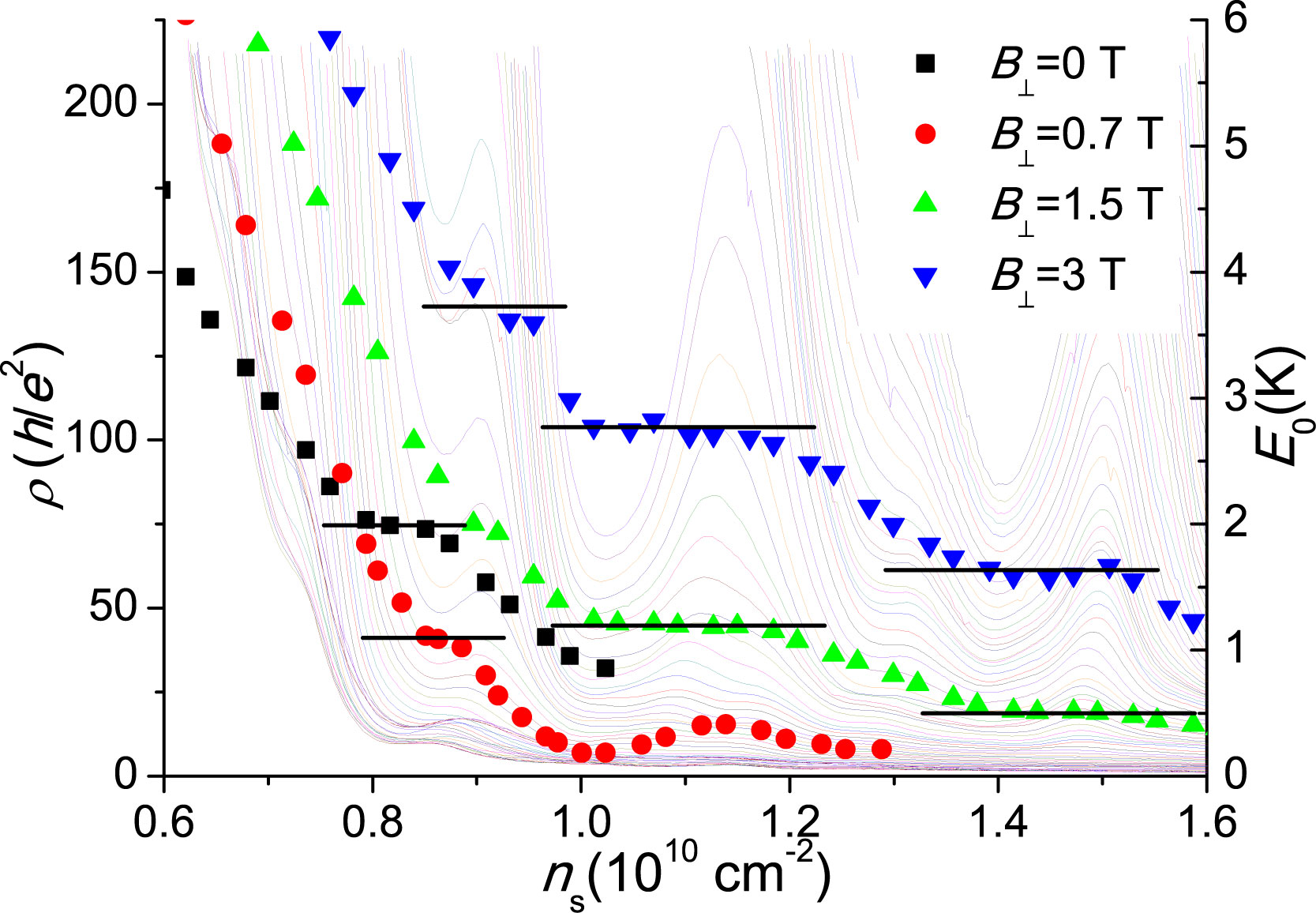}\newline%Rb13
\caption{(Colour online). Activation energies at various magnetic fields obtained from the high temperature activated behaviour as a function of electron density in comparison to the magnetically induced resistance oscillations (A78a). On an overall decrease with increasing $n_{\rm{s}}$, clear plateaux are formed, starting approximately at the position of a trough and ending at the next peak.}
\label{fig3}
\end{figure}

\begin{figure}[h]
\centering
\includegraphics[width=0.45\textwidth]{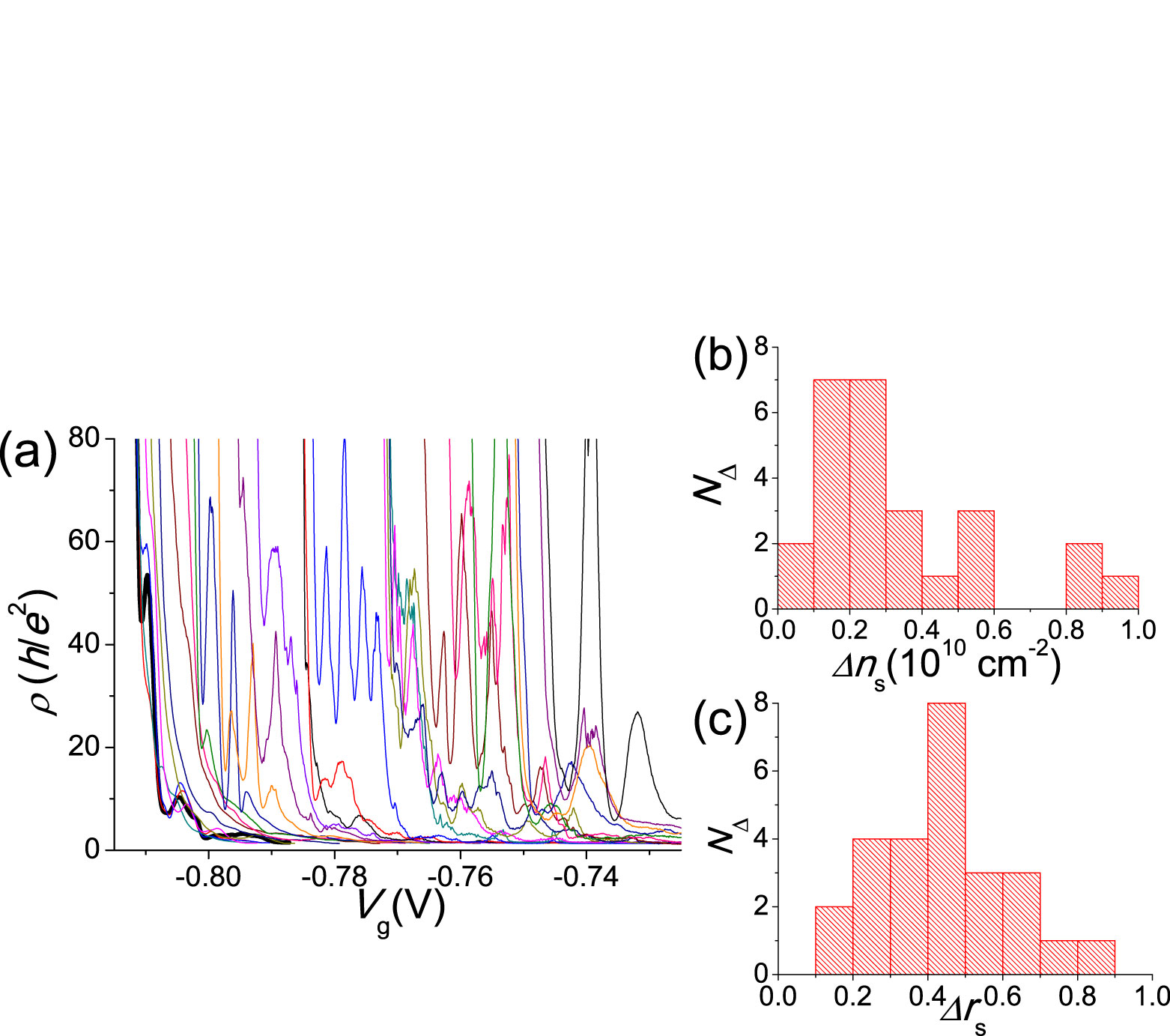}%PeakHistogram
\caption{(Colour online).(a) $\rho$ vs. $V_{\rm{g}}$ of device A78b at stepwise increasing $B_{\perp}=0-2.75$\,T ($\Delta B_{\perp}=0.1$\,T), after a ``bad'' (fast) cooldown. (b), (c) Histograms of the separation between adjacent peaks in terms of $n_{\rm{s}}$ and $r_{\rm{s}}$. The data includes 10 devices from various wafers and of varying dimensions.}
\label{fig4}
\end{figure}

The activated part of the $T$ dependence provides important information about the energy scale that dominates transport in this regime, via the activation energy $E_0$ which can be extracted from fits of the form $\rho(T)=\rho_0 \exp(E_0/k_{\rm{B}}T)$ (solid lines in the lower panels of Fig.~\ref{fig2}), with $k_{\rm{B}}$ the Boltzmann constant. In Fig.~\ref{fig3} $E_0$ is plotted as a function of $n_{\rm{s}}$ for various $B_{\perp}$, combined with the resistance oscillation of the same device.
It exhibits a striking behaviour: An overall decrease of $E_0$ with increasing $n_{\rm{s}}$ is observed, but clear plateaux are formed between troughs and next following peaks. The plateaux (highlighted by horizontal lines) are strongest at higher fields $B_{\perp}=1.5$ and 3\,T, but are observed even at $B_{\perp}=0$, where resistance oscillations are absent. This behaviour is particularly remarkable when noticing that the exponential fits are done in the high $T$ regime where the oscillations are suppressed in simple resistance measurements. If the activated behaviour is caused by excitation of charge carriers to a mobility edge (e.g. activation across a potential barrier), the $n_{\rm{s}}$ dependence of $E_0$ provides information about the compressibility of the system, with a flat behaviour evidence of a pinning of the Fermi energy $E_{\rm{F}}$ by a highly compressible phase. In this view, the observed behaviour would suggest a repeated transition between compressible and incompressible states. Note that the monotonic $n_{\rm{s}}$ dependence of $E_0$ also gives further evidence against a CB effect, where an oscillatory $V_{\rm{g}}$ dependence of $E_0$ is expected, with $E_{0}$ disappearing at resonance (resistance minima).

The large separation in terms of $n_{\rm{s}}$ between the peaks suggests a phenomenon involving many electrons as the origin of the observation, rather than single charging events. We have previously presented evidence of formation of a disorder stabilised electron solid (ES) in our systems, where transport occurs by tunnelling of defects. This was mainly supported by the observation that the average hopping distance of the charge carriers was approximately equal to the average electron-electron separation~\cite{Ghosh2004,Baenninger2005}. If a pinned ES was formed in our relatively disordered 2DES, one would expect a complex interplay between the ES (or, indeed, any strongly correlated electron phase) and the background disorder potential as the electron density is varied. This could lead to a "locking" of the ES over a certain density range, where most of the electrons remain in their positions while additional electrons are added to interstitial positions. At some point a large scale rearrangement/relaxation of the solid may occur with a strong impact on transport properties. Such repeated rearrangements, possibly accompanied by changes in the degree of ordering in the solid, could explain the the resistance oscillations, as well as the behaviour of $E_0$ (plateaux when the solid is "locked" with a pinning of $E_{\rm{F}}$ due to a large density of defect states). The peculiar $T$ dependence could then arise from a transition from activated transport (high $T$) to a near-resonant tunnelling of defects (low $T$) with metallic behaviour caused by defect delocalisation if disorder is not too strong~\cite{Baenninger2008}. The magnetic field may be required to induce or stabilise the ES by reducing the wave function overlap between electrons, an effect that has long been proposed~\cite{Lozovik1975}. Another possible mechanism based on stripe formation in an electron lattice has been proposed to cause resistance oscillations~\cite{Slutskin2000}.

Finally, we will discuss another unexpected observation, namely that the positions of the peaks may not be entirely random. It was found that most devices showed particularly strong peaks in the proximity of $n_{\rm{s}}\approx$ 1.2 and 1.8$\times 10^{10}$\,cm$^{-2}$. In view of previous results highlighting the importance of the average electron-electron separation $r_{\rm{ee}}\approx 1/\sqrt{n_{\rm{s}}}$~\cite{Ghosh2004,Baenninger2005} we have analysed the peak positions in terms of the interaction parameter $r_{\rm s}=1/a_{\rm B}^{*}\sqrt{\pi n_{\rm s}}$ ($a_{\rm B}^{*}$ the effective Bohr radius), which gives a dimensionless expression for $r_{\rm{ee}}$. One immediately notices that $n_{\rm{s}}\approx 1.2$ and $1.8\times 10^{10}$\,cm$^{-2}$, correspond to the integer values $r_{\rm{s}}\approx 5$ and 4. Furthermore, we found the separation between peaks to be approximately 0.5 in terms of $r_{\rm{s}}$. This is confirmed by the histograms of the peak separation shown in Figs.~\ref{fig4} (b), and (c), where we have anylised 10 devices from five different wafers ($\delta$-doped with $\delta_{\rm{sp}}=20$, 40, 60 and 80\,nm and bulk doped with $\delta_{\rm{sp}}=40$\,nm with $W\times L=8\,\mu\rm{m}\times 0.5-4\,\mu\rm{m}$). While (b) exhibits a peak around $\Delta n_{\rm{s}}\approx 0.2\times10^{10}$\,cm$^{-2}$ but otherwise apparently random distribution (c) shows a clear, if broad maximum around $\Delta r_{\rm{s}}\approx 0.4-0.5$. A similar analysis in terms of absolute value of $r_{\rm{s}}$ is difficult because of the relatively small number of devices and the experimental error in $n_{\rm{s}}$. Nevertheless, our observation gives some indication of a distinction of integer and half integer values of $r_{\rm{s}}$, which is further supported by previous experiments in the localised regime of 2DES of similar dimensions to the ones used here~\cite{Pepper1979,Ghosh2004a}. Although a definite statement in this matter will require more work, a possible universality in such a fundamental quantity as $r_{\rm{s}}$ should not go unmentioned due to its far reaching implications. It would suggest fundamental density dependent instabilities of currently unknown origin, which could not be directly related to details of background disorder, although disorder could still play a role in enhancing electron-electron interactions in combination with the magnetic field.

We thank EPSRC and the UK-India Education and Research Initiative (UKIERI) for financial support. MB thanks the COLLECT network and the Sunburst and Cambridge Overseas Trusts.

%\bibliography{OscRefs}

\end{document}